\documentclass[aps,a4paper,twocolumn]{revtex4}

\usepackage{amsmath}
\usepackage{graphicx}
\usepackage{subfigure}
\usepackage[usenames,dvipsnames]{color}
\definecolor{darkblue}{RGB}{0,0,196}
\usepackage[colorlinks=true,linkcolor=darkblue,citecolor=darkblue,urlcolor=darkblue]{hyperref}
\usepackage{setspace}
\usepackage{multirow}
\usepackage{hyperref}
\usepackage{xcolor}
\usepackage{xspace}
\hypersetup{
  colorlinks   = true, 
  urlcolor     = blue, 
  linkcolor    = blue, 
  citecolor   = blue 
}
\usepackage{graphicx}
\usepackage{amsmath,bbm}
\usepackage{amssymb,bm}
\usepackage{yfonts}
\usepackage{comment}
\usepackage[normalem]{ulem}

\newcommand{\pT}{\ensuremath{p_{\rm T}}\xspace}

\newcommand{\Mpk}{\ensuremath{M_{\rm pK}}\xspace}
\newcommand{\Mdk}{\ensuremath{M_{\rm (d/2)K}}\xspace}

\begin{document}

\title{$\Lambda$(1520) as a probe of resonance-driven deuteron formation at the LHC}
\author{Sushanta Tripathy\footnote{sushanta.tripathy@cern.ch}}
\author{Peter Christiansen\footnote{peter.christiansen@cern.ch}}

\affiliation{Division of Particle and Nuclear Physics, Department of Physics, Lund University, Lund, 22364, Sweden}

\date{\today}

\begin{abstract}
Light nuclei such as deuterons are produced abundantly in high-energy proton-proton and nuclear collisions despite their tiny binding energies. Their production mechanism remains unresolved, as both nucleon coalescence and statistical thermal models reproduce inclusive LHC yields. We propose a direct invariant-mass observable that discriminates between these scenarios using the long-lived $\Lambda(1520) \to {\rm pK}$ resonance. If decay protons coalesce into deuterons, the produced nuclei remain correlated with the kaon, allowing the resonance peak to be reconstructed experimentally through proxy masses, \Mdk, formed from kaons and half the deuteron four-momentum. Using Thermal-FIST and PYTHIA with a deuteron coalescence afterburner, we show that a \Mdk peak emerges only in the coalescence scenario. This observable provides a direct experimental probe of resonance-fed deuteron production and of late-stage coalescence dynamics in high-energy collisions.

\end{abstract}
  
\maketitle



In ultra-relativistic collisions at the LHC at CERN, it remains challenging to understand how loosely bound objects such as the deuteron, with binding energy of about 2.2 MeV, emerge from a system with characteristic hadronic scales of order 100 MeV. Traditionally, light-nuclei production is modeled via late-stage coalescence of nucleons once the system has cooled and diluted ~\cite{Sato:1981ez,Nagle:1996vp,Scheibl:1998tk,Blum:2017qnn,Blum:2019suo,Mrowczynski:2019yrr,Bellini:2020cbj,Mahlein:2023fmx,Korsmeier:2017xzj,Kachelriess:2020uoh,vonDoetinchem:2020vbj,PhysRevD.105.083021}. However, the progress in statistical-thermal modeling, in which hadrons and nuclei are emitted from a common source in approximate chemical equilibrium, have revealed that these models can also reproduce the inclusive light-nuclei yields at the LHC with remarkable accuracy ~\cite{SHM1,SHM2,SHM3,SHM4,SHM5,SHM6}. Since these contrasting pictures describe the same inclusive observables, it raises questions about the nature of the light-nuclei production mechanism~\cite{Tripathy:2025hly, BraunMunzinger:2019, Bellini:2019, Mrowczynski:2020ugu}. This puzzle also has consequences for some indirect dark-matter searches, where the secondary production of light antinuclei in hadronic interactions constitutes an irreducible background~\cite{Korsmeier:2017xzj,vonDoetinchem:2020vbj,PhysRevD.105.083021}. 

\begin{figure*}
    \centering
    \includegraphics[width=0.8\linewidth]{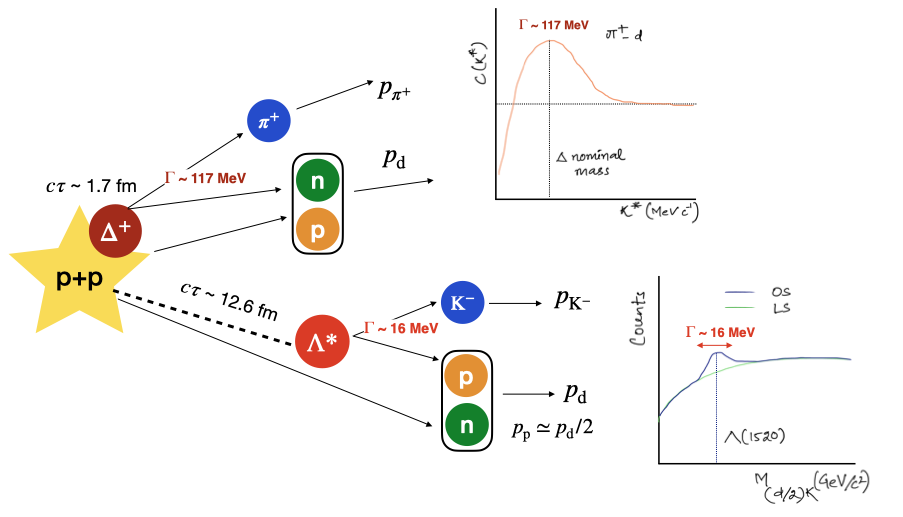}
    \caption{Illustration of the proposed observable and its complementarity to existing femtoscopic measurements. For short-lived resonances, as observed by ALICE Collaboration~\cite{ALICE:2025byl}, coalescence occurs close to the collision point and the large resonance width induces significant momentum smearing. In contrast, the long-lived resonances studied here decay far from the production point, while their narrow widths permit a clean reconstruction of the resonance origin. The reduced smearing may additionally constrain the coalescence dynamics itself, cf.\ Fig.~\ref{fig:p0r0test}. The illustrated proton-neutron configuration is schematic; no restriction on the production origin of the nucleons is imposed in the inclusive coalescence calculation.}
    \label{fig:illustration}
\end{figure*}

A recent ALICE femtoscopic analysis has provided first evidence for resonance-fed deuteron formation, consistent with coalescence~\cite{ALICE:2025byl}. By measuring pion-deuteron correlations in high-multiplicity pp collisions at $\sqrt{s}=13$ TeV, ALICE isolated the imprint of short-lived baryon resonances, mainly $\Delta(1232)$ baryons, on the ${\rm \pi}$–d correlation function and concluded that a substantial fraction of deuterons is associated with nucleons emitted in strong resonance decays. 
While this provides evidence that resonance feed-down is important, the short lifetime of the baryon resonances, which is of the order of the size and lifetime of the system formed in proton-proton collisions, means that one is still in a rather dense environment. Here, we instead study the longer-lived $\Lambda(1520)$ (also referred to as $\Lambda^*$), with an average lifetime of approximately 13 fm/c. While significant suppression of the $\Lambda(1520)$ has been observed in central Pb–Pb collisions~\cite{ALICE:2018ewo}, measurements in pp and p–Pb collisions show no evidence for a comparable suppression in small collision systems~\cite{Acharya:2019yoi}. Its longer lifetime therefore makes the $\Lambda(1520)$ a suitable probe of nucleon coalescence at later times and in a more dilute environment.
Our idea is focused on exploring the main decay channel, $\Lambda(1520)\rightarrow pK$, which is experimentally easy to identify due to its relatively narrow  width, $\Gamma \simeq 15.7$ MeV~\cite{ParticleDataGroup:2024}: 
 \begin{equation}
     \Mpk^2 = (p_{\rm p} + p_{\rm K})^2,
 \end{equation}
where $p_{\rm p}$ and $p_{\rm K}$ are the proton and kaon four-momentum vectors. To identify deuterons that are formed via coalescence of the proton from $\Lambda(1520)$ decays, we therefore construct a proxy mass \Mdk using a kaon and a proxy proton four momentum $p_{\rm proxy}\simeq p_{\rm d}/2$:
\begin{equation}
     \Mdk^2 = \bigg(\frac{p_{\rm d}}{2} + p_{\rm K}\bigg)^2.
\end{equation}
If the deuteron sample contains a contribution from $\Lambda(1520)$ protons and the coalescence process does not introduce too large smearing, a $\Lambda(1520)$-like resonance peak should appear in the  \Mdk spectrum; otherwise the \Mdk spectrum is expected to remain smooth after background subtraction. 
While $\Delta(1232)$ is expected to be the dominant resonance source of decay nucleons relevant for deuteron formation, its much shorter lifetime and correspondingly larger natural width, together with the available decay phase space, make it less suitable for the proposed invariant-mass approach. In contrast, the long lifetime and narrow width of the $\Lambda(1520)$ provide both a cleaner invariant-mass signature and access to deuteron formation in a more dilute environment.
The proposed measurement is illustrated in Fig.~\ref{fig:illustration}.


To demonstrate the capabilities of the new measurement to separate between different production mechanisms, we compare two contrasting model baselines. For both models, we study the standard \Mpk spectrum and the proxy \Mdk spectrum in the same event class and kinematic range. Thermal-FIST (Thermal, Fast and Interactive Statistical Toolkit)~\cite{Vovchenko:2019pjl} is a statistical hadronization model based on particle production under the assumption of thermal equilibrium. It is based on the hadron resonance gas picture, in which particle yields and fluctuations are determined by very few thermodynamic parameters such as temperature, chemical potentials, and correlation volumes. In this framework, QCD enters effectively through the spectrum of hadronic states included in the calculation. To obtain more realistic kinematics, the generated particles are boosted with a Blast-Wave parameterization tuned to ALICE pp data at $\sqrt{s}=$ 13 TeV~\cite{ALICE:2020nkc}, that the resulting $p_{\rm T}$ spectra are comparable to the measured ones. This setup has previously been used as a statistical baseline for ALICE measurements of light-(anti)nuclei production~\cite{ALICE:2020foi,ALICE:2022veq, Vovchenko:2018fiy}, as well as for the extraction of chemical potentials from identified particle yield ratios in Pb--Pb collisions~\cite{ALICE:2023ulv}.
In contrast, for coalescence, we use PYTHIA 8.3~\cite{Bierlich:2022pfr} supplemented with a deuteron coalescence afterburner. PYTHIA is the most widely used Monte Carlo event generator~\cite{Bierlich:2022pfr} at LHC and inspired by perturbative QCD for partonic collisions, combined with the Lund string model for hadronization. In this framework, color fields between partons are represented as strings~\cite{Andersson:1983ia}, which fragment into hadrons through successive string breakings. It is then followed by an event-by-event coalescence afterburner which produces deuterons from nucleon pairs in the pair rest frame. This setup has previously been used successfully to describe the transverse-momentum spectra and coalescence parameter of (anti-)deuterons measured by ALICE in high-multiplicity pp collisions at $\sqrt{s}=13$ TeV, with the best agreement obtained  using the Argonne $v_{18}$ deuteron wave function~\cite{Mahlein:2023fmx}.

\begin{figure*}[ht!]
    \centering
\includegraphics[width=0.85\linewidth]{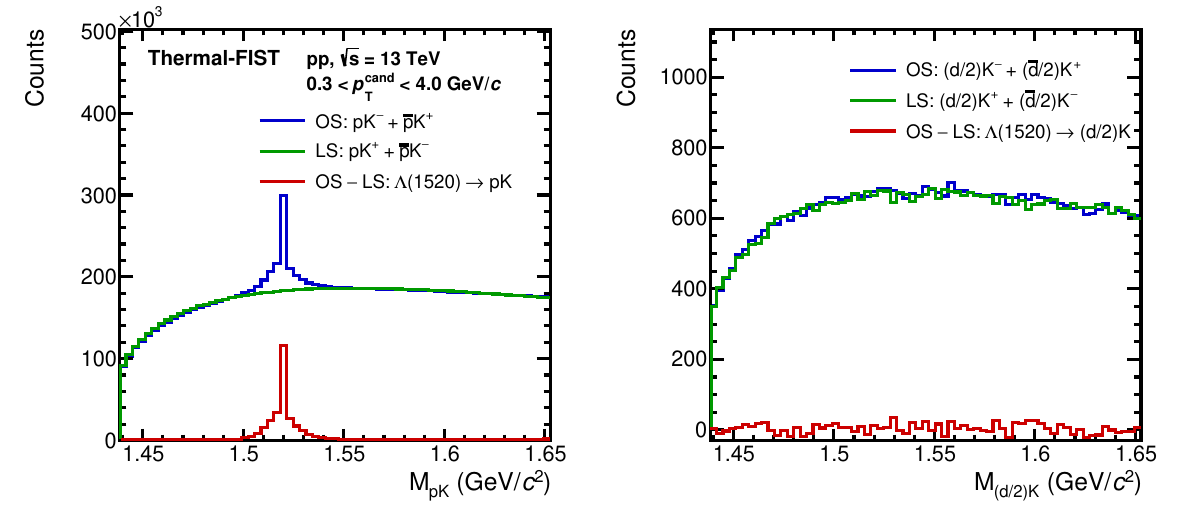}
    \caption{$\Lambda$(1520) candidate invariant mass distributions for the pK channel (left) and the (d/2)K proxy channel (right), shown for 0.3–4.0 GeV/c candidate-\pT range in pp collisions at $\sqrt{s}$ = 13 TeV obtained using Thermal-FIST model. Here, both the opposite sign (OS) and like sign (LS), and the background subtracted signal (OS--LS) distributions are shown.}
\label{fig:FIST}
\end{figure*}

For the present study, deuteron formation is modelled using the Argonne v18 wave function in the coalescence afterburner, following an implementation similar to that developed in Ref.~\cite{Mahlein:2023fmx}. All protons and neutrons in each event are considered for coalescence irrespective of their production origin, thereby including both prompt and resonance-decay nucleons. For each proton, the coalescence probability with the available neutrons is evaluated from their relative spatial separation and momentum in the pair rest frame using the Argonne v18 wave function. The neutron giving the largest coalescence probability is selected as the candidate partner and the pair is accepted probabilistically according to this weight. Once a pair is accepted, both nucleons are excluded from subsequent combinations. The same procedure is applied independently to antinucleons. This approach has been successfully used in the interpretation of the previously discussed ALICE femtoscopic analysis, which provided the first evidence for resonance-fed deuteron production~\cite{ALICE:2025byl}. The production space-time coordinates of the nucleons are taken from PYTHIA. For each proton-neutron pair, the four-momenta and production coordinates are transformed to the pair rest frame, where the earlier-produced nucleon is propagated along its trajectory to the production time of the later nucleon. The coalescence probability is then evaluated using the resulting equal-time spatial separation $r^*=|\mathbf{x}_p^*-\mathbf{x}_n^*|$ and relative momentum $k^*=|\mathbf{p}_p^*-\mathbf{p}_n^*|/2$, together with the Argonne v18 wave function. The spatial separation is determined event-by-event from the generated production vertices. Since PYTHIA does not natively provide the state required for the proxy analysis, a modification proposed by PYTHIA authors was used to introduce a $\Lambda(1520)$-like resonance at hadronization by promoting a fraction of produced $\Lambda$- and $\Sigma$-like baryons into $\Lambda(1520)$ states. To make the PYTHIA results comparable with Thermal-FIST, the fractions of primary and resonance-fed accepted protons in PYTHIA are reweighted to reproduce the corresponding FIST expectations. In addition, we constrain the input production baseline of both models to reproduce the measured values of proton and $\Lambda(1520)$ integrated yields in the corresponding kinematic ranges for pp collisions at $\sqrt{s}$ = 13 TeV~\cite{Padhan:2022Lambda1520,ALICE:2020jsh}. This is implemented through separate normalization factors applied to the inclusive proton and $\Lambda(1520)$ contributions at the input level. Importantly, these factors are used only to fix the integrated production rates and do not modify the PYTHIA-generated kinematic distributions. In this way, the obtained distributions for the $\Lambda(1520)\to {\rm pK}$ channel and the $\Lambda(1520)\to {\rm (d/2)K}$ proxy channel are expected to be as realistic as possible. 

To make observable predictions, the \Mpk and \Mdk spectra are studied in kinematic ranges where the ALICE experiment has acceptance, $|\eta| < 0.8$, and is expected to be able to do the particle identification with high purity and efficiency, $0.3 < p_{\rm T}^{\rm cand} < 4.0$ GeV/$c$, where $p_{\rm T}^{\rm cand}$ is the transverse momentum of the resonance candidate. To isolate the resonance peak, the combinatorial background is estimated from the like-sign (LS) invariant mass spectra, i.e., while the \Mpk signal is obtained using opposite-sign (OS) pairs of ${\rm pK}^-$ and ${\bar{\rm p}\rm K}^+$, the \Mpk background is found from LS pairs of ${\rm pK}^+$ and ${\bar{\rm p}\rm K}^-$.

Figure~\ref{fig:FIST} shows results from 1.1 Billion events of Thermal-FIST. While the \Mpk spectrum shows a clear $\Lambda(1520)$ peak, the \Mdk spectrum does not develop a resonance component. This is the expected behavior if deuterons are not preferentially coalesced from resonance-decay nucleons, and Thermal-FIST therefore provides a useful null signal reference for the proxy observable.

\begin{figure*}
    \centering
    \includegraphics[width=0.45\linewidth]{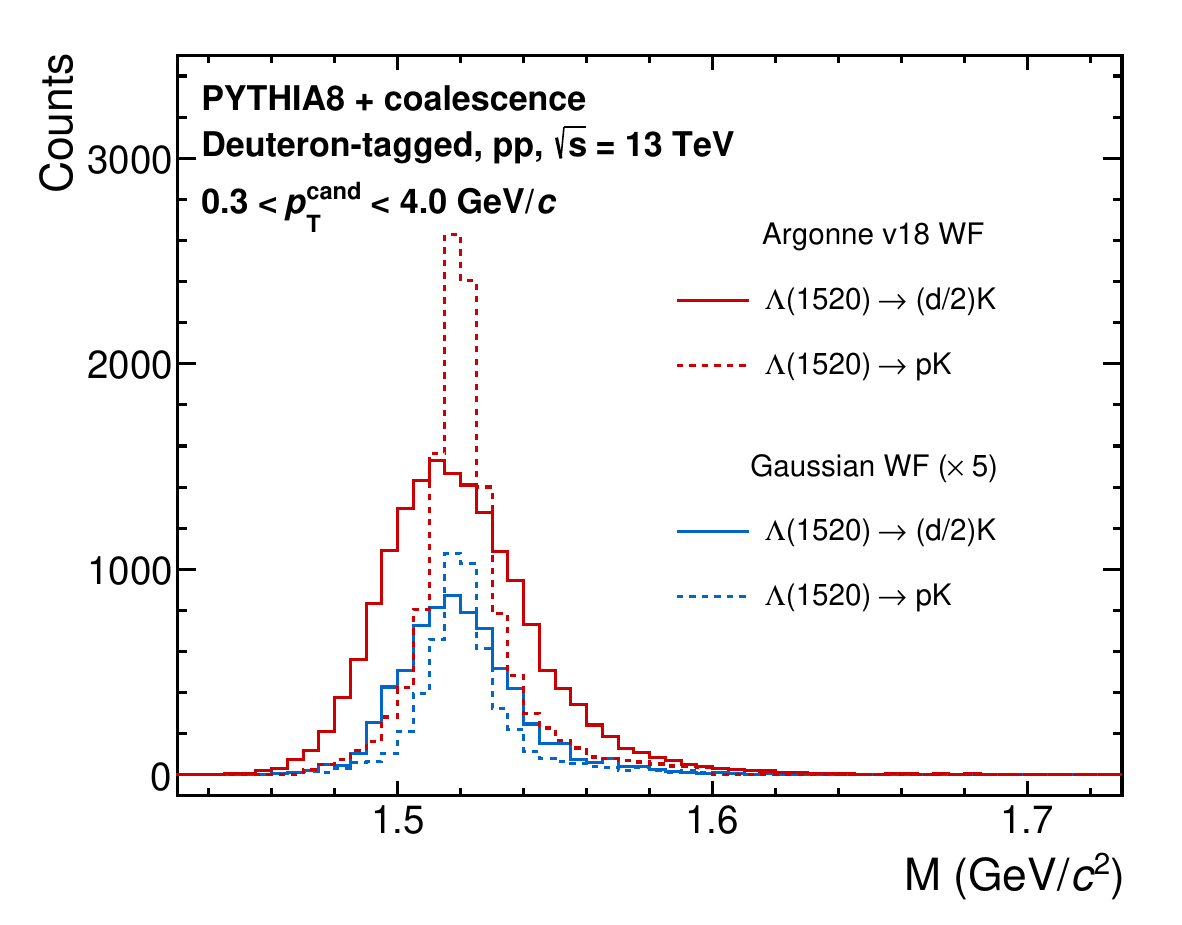}
        \includegraphics[width=0.45\linewidth]{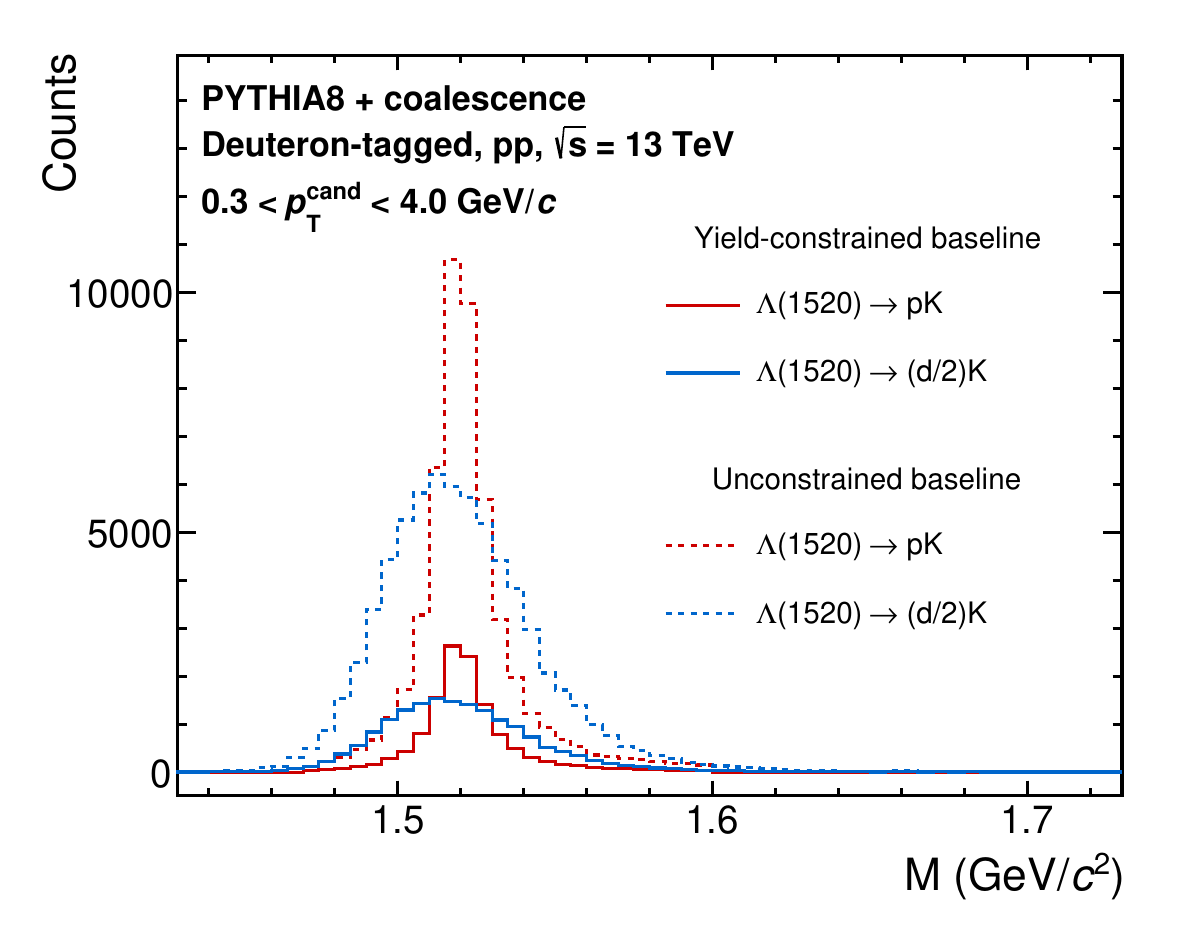}
    \caption{Deuteron-tagged $\Lambda(1520)$ signal distributions in PYTHIA with coalescence afterburner for pp collisions at $\sqrt{s}=13$ TeV, shown for $0.3 < p_{\rm T}^{\rm cand} < 4.0$ GeV/$c$. Left: $\Lambda(1520)\to{\rm (d/2)K}$ with the corresponding $\Lambda(1520)\to{\rm pK}$ reference for two choices of the deuteron wave function, Argonne v18 and Gaussian. The Gaussian wave function results are scaled by a factor of 5 for better visibility. Right: same two channels using the Argonne v18 wave function, comparing the yield-constrained baseline with the unconstrained baseline. In all cases, the proton entering the deuteron is required to originate from the generated $\Lambda(1520)$, and the kaon is required to be the daughter from the same $\Lambda(1520)$ decay.}
    \label{fig:PYTHIA8_taggedOverlay}
\end{figure*}

\begin{figure*}
    \centering
    \includegraphics[width=0.45\linewidth]{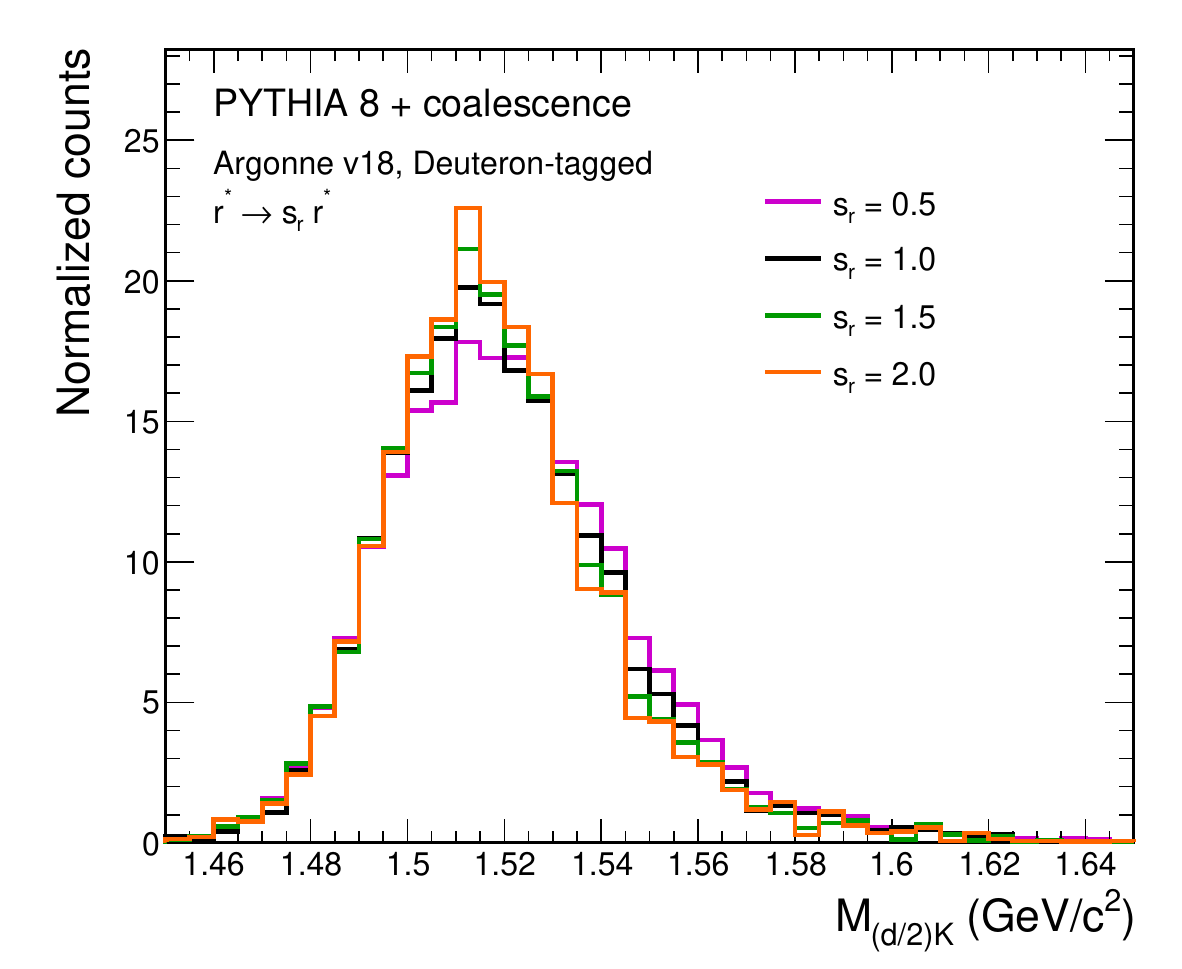}
        \includegraphics[width=0.45\linewidth]{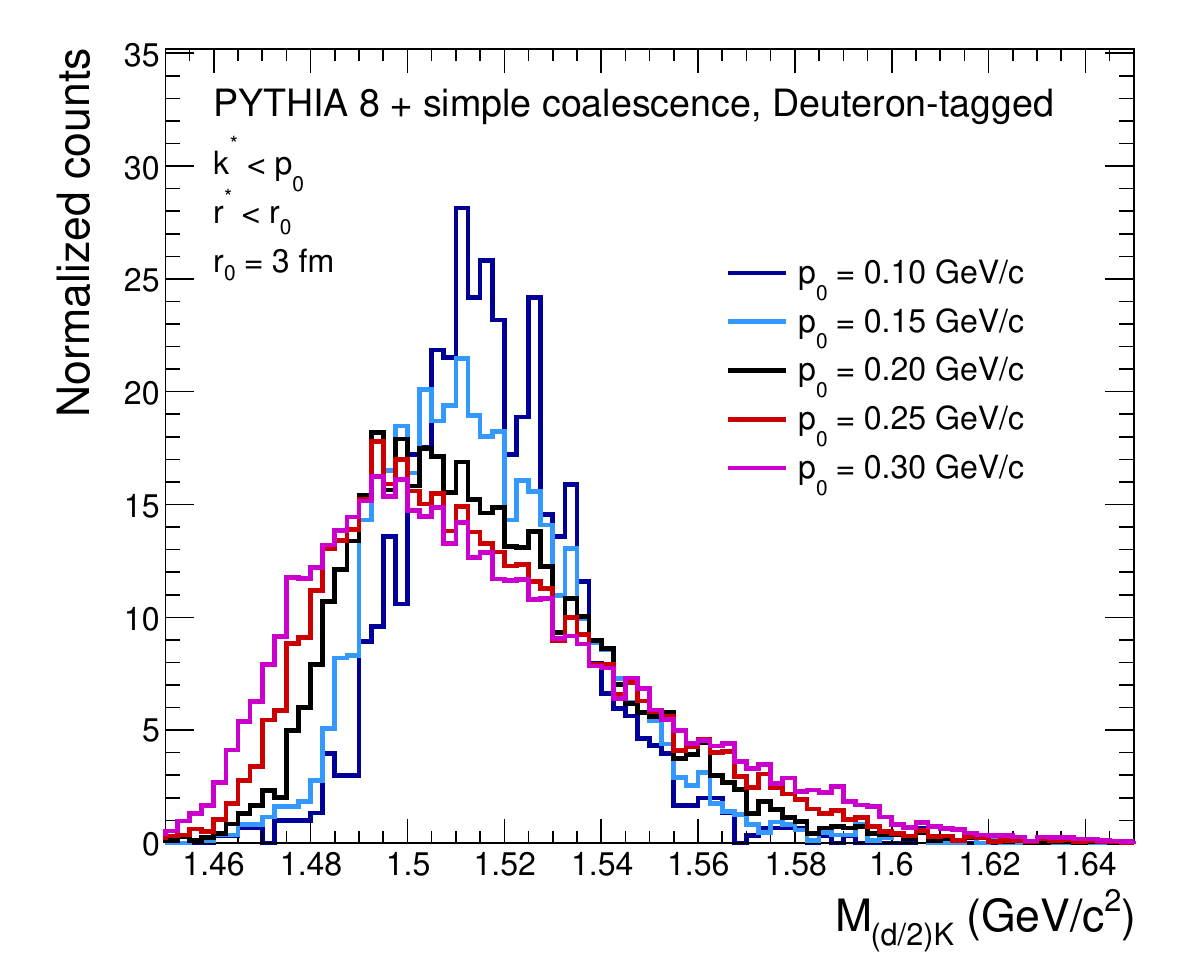}
    \caption{Sensitivity of the deuteron-tagged $\Lambda(1520)\rightarrow {\rm (d/2)K}$ proxy shape to the coalescence conditions in PYTHIA 8. Left: normalized $M_{\rm (d/2)K}$ distributions obtained with the Argonne v18 wave function after rescaling the relative spatial separation as $r^*\rightarrow s_r r^*$. Right: normalized $M_{\rm (d/2)K}$ distributions obtained with a simplified phase-space coalescence prescription, $k^*<p_0$ and $r^*<r_0$, for different values of $p_0$ at fixed $r_0$ = 3 fm. The distributions are normalized to the same integral in the considered signal region to emphasize shape modifications.}
    \label{fig:p0r0test}
\end{figure*}

The discriminating power of the observable becomes apparent when contrasted with the PYTHIA-based coalescence framework introduced above. For this study, we generate nearly 1.5 Billion minimum bias events, which corresponds to only about 0.01\% of the total events recorded by ALICE experiment in Run 3 so far ~\cite{Pinto2026SQM}.  We first use the full MC information to test the robustness of the new observable, shown in Fig.~\ref{fig:PYTHIA8_taggedOverlay}, by demonstrating that the proxy preserves the parent-resonance structure in all cases. In these tests, the proton entering the deuteron is required to originate from a generated $\Lambda(1520)$, and the kaon is required to be a daughter of the same $\Lambda(1520)$ decay. This construction removes combinatorial ambiguities and directly tests whether the reconstructed $\Lambda(1520)\to{\rm (d/2)K}$ proxy remains correlated with the corresponding $\Lambda(1520)\to{\rm pK}$ reference. The comparison is performed for both the Argonne v18 and Gaussian deuteron wave functions, and also with and without the yield-constraining normalization factors applied to the input proton and $\Lambda(1520)$ production rates. While the absolute yields change, as expected from the different coalescence probabilities and input production normalizations, the proxy signal consistently exhibits a peak in the same mass region as the truth-matched $\Lambda(1520)\to{\rm pK}$ reference. This demonstrates that the appearance of the proxy signal is not generated by the reweighting procedure, but follows from the kinematic connection between the coalesced deuteron and the original proton from the $\Lambda(1520)$ decay.  Beyond establishing the presence of the proxy signal, its broadening relative to the conventional $\Lambda(1520)\rightarrow pK$ peak can itself provide information on the coalescence conditions. To illustrate this sensitivity, Fig.~\ref{fig:p0r0test} explores separately variations of the spatial and momentum components entering the coalescence condition. The left panel of Fig.~\ref{fig:p0r0test} shows the effect of rescaling the relative spatial separation entering the Argonne v18 coalescence calculation, $r^*\rightarrow s_{\rm source}r^*$. Over the considered range, the normalized proxy distribution exhibits only moderate variations. In contrast, the right panel shows a simplified phase-space coalescence prescription, $r^*<r_0$ and $k^*<p_0$, at fixed $r_0$=3 fm. Varying $p_0$ from 0.10 to 0.30 GeV/c produces a pronounced modification of the width and shape of the proxy signal. This demonstrates that, in addition to identifying resonance-fed deuteron formation, the proxy shape retains sensitivity to the phase-space conditions governing coalescence, particularly to the relative momentum of the coalescing nucleons.

\begin{figure*}[ht!]
    \centering
\includegraphics[width=0.85\linewidth]{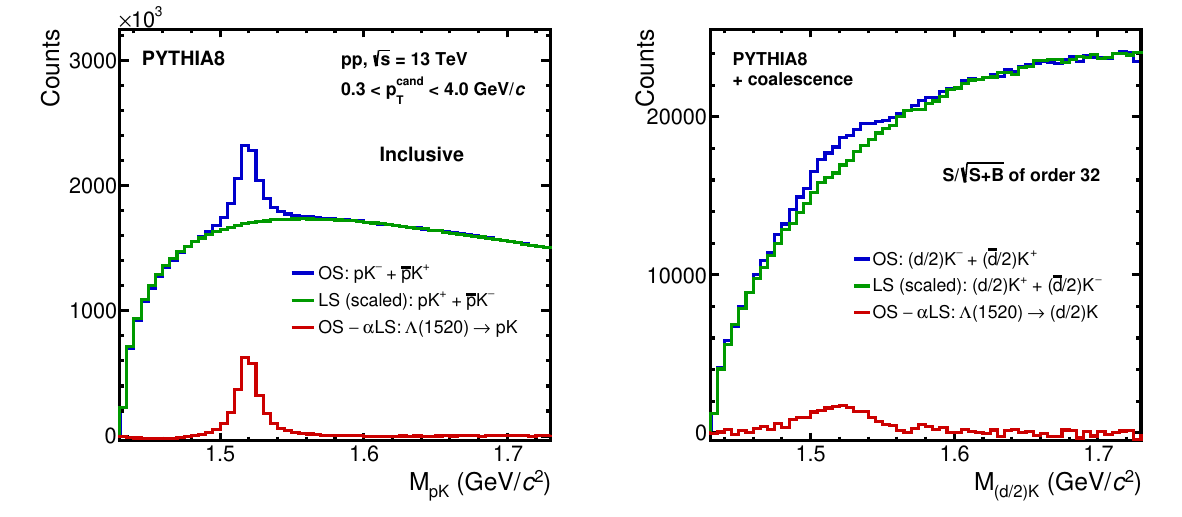}
    \caption{$\Lambda$(1520) candidate invariant mass distributions for the pK channel (left) and the (d/2)K proxy channel (right), shown for 0.3–4.0 GeV/c candidate-\pT range in pp collisions at $\sqrt{s}$ = 13 TeV obtained using PYTHIA with coalescence afterburner. Here, both the opposite sign (OS) and like sign (LS), and the background subtracted signal (OS--$\alpha$LS) distributions are shown. For the inclusive proxy channel, the signal yield gives a statistical significance of $\rm S/\sqrt{S+B}\simeq32$ evaluated in the window $m_{\Lambda(1520)}\pm3\Gamma$.}
\label{fig:PYTHIA8}
\end{figure*}

Having established the robustness of the new observable under ideal conditions, we then turn to the inclusive case shown in Fig.~\ref{fig:PYTHIA8} which shows the PYTHIA plus coalescence predictions for the \Mpk and \Mdk spectra. Here, no requirement is imposed on the origin of the deuteron. The top panel shows the inclusive opposite-sign (OS), like-sign (LS), and background-subtracted invariant-mass distributions in the interval $0.3<p_{\rm T}^{\rm cand}<4.0~{\rm GeV}/c$. A clear $\Lambda(1520)$ peak is seen in the standard pK reconstruction. Importantly, after subtraction of the large combinatorial background, a visible peak also emerges in the inclusive proxy channel in the same mass region. The signal distributions are extracted from the background-subtracted combination $\rm S(M)=OS(M)-\alpha\,LS(M)$, where the normalization factor $\alpha$ is obtained by matching the LS spectrum to the OS spectrum in a high-mass sideband $1.60 < M < 1.70~{\rm GeV}/c^2$, well separated from the signal region. The stability of the background normalization was tested by varying the sideband to $1.60 < M < 1.65$ and $1.65 < M < 1.70$ GeV/$c^2$.  The extracted proxy yield changes by at most 5.3\%, which is assigned as a systematic uncertainty associated with the background normalization. Thus, even in the inclusive case, the proxy retains sensitivity to the underlying $\Lambda(1520)$ structure. Using the window $m_{\Lambda(1520)} \pm 3\Gamma$, corresponding to $1.47 < M_{\rm (d/2)K} < 1.57~{\rm GeV}/c^2$, the extracted proxy yield is $S = 1.88\times 10^4$ over a scaled like-sign background of $B = 3.28\times 10^5$, giving a statistical significance $S/\sqrt{S+B} \simeq 32$. Given the shown simulated sample corresponds to less than 0.01\% of the ALICE Run~3 statistics discussed above, the statistical sensitivity of the proxy channel is therefore expected to be more than sufficient. The experimental challenge is therefore not the statistical extraction of the peak given large pp data samples already collected at the LHC, instead the control of experimental uncertainties associated with particle-identification purity, deuteron reconstruction efficiency, and the stability of the combinatorial-background subtraction will be crucial.

In this Letter, we have shown that $\Lambda(1520)\to {\rm pK}$ provides a direct and experimentally accessible probe of deuteron coalescence. Although $\Lambda(1520)$ decays contribute only about 0.8\% of the accepted final-state proton yield in the Thermal-FIST baseline, their reconstructible pK decay and long lifetime make this channel a uniquely clean benchmark for studying nucleus formation in a dilute environment far from the collision point. Within the PYTHIA plus coalescence framework, the proposed proxy observable directly tests whether nucleons originating from long-lived resonance decays can subsequently participate in deuteron formation. More generally, the observable maps the microscopic parentage of deuterons onto a measurable invariant-mass signature, opening a new avenue for experimentally constraining the space-time dynamics of light-nucleus production in high-energy collisions. With the large pp data samples already collected by the LHC experiments during Run 3, such measurements appear feasible in the near future.

\section*{Acknowledgment}
The authors acknowledge the funding received from the European Union’s Horizon Europe research and innovation programme under the Marie Skłodowska-Curie grant agreement No. 101149298. The authors thank T.~Sjöstrand and C.~Bierlich for help with implementing the $\Lambda(1520)$ production in PYTHIA.  

\bibliography{References}



\end{document}